\magnification\magstephalf
\tolerance 10000
\input epsf.tex
\def\tr{{\rm tr}}
\font\rfont=cmr10 at 10 true pt
\def\ref#1{$^{\hbox{\rfont {[#1]}}}$}
\def\k{{\bf k}}\def\x{{\bf x}}
\def\M{{\bf M}} \def\P{{\pmb{$\Pi$}}}\def\D{{\bf D}}
\font\fig=cmr10 at 10truept


\font\fourteenbf=cmbx12 scaled\magstep1

 \def\b {{\beta }}  
\def\e{\epsilon}  \def\o{\omega}

\def\pd {\partial}
\def\pmb#1{\setbox0=\hbox{#1}
 \kern.05em\copy0\kern-\wd0 \kern-.025em\raise.0433em\box0 }

\def\slash{/\kern-.5em}

\def \half {{\scriptstyle {1 \over 2}}}

 %


\def\boxit#1{\vbox{\hrule\hbox{\vrule\kern1pt\vbox
{\kern1pt#1\kern1pt}\kern1pt\vrule}\hrule}}

\def\h{\hfill\break}
\parskip=6pt
\parindent=0pt
\hsize=17truecm\hoffset=-5truemm
\voffset=-1truecm\vsize=24.5truecm
\def\footnoterule{\kern-3pt
\hrule width 17truecm \kern 2.6pt}


\catcode`\@=11 

\def\nolabels{\def\wrlabeL##1{}\def\eqlabeL##1{}\def\reflabeL##1{}}
\def\writelabels{\def\wrlabeL##1{\leavevmode\vadjust{\rlap{\smash%
{\line{{\escapechar=` \hfill\rlap{\sevenrm\hskip.03in\string##1}}}}}}}%
\def\eqlabeL##1{{\escapechar-1\rlap{\sevenrm\hskip.05in\string##1}}}%
\def\reflabeL##1{\noexpand\llap{\noexpand\sevenrm\string\string\string##1}}}
\nolabels
\global\newcount\refno \global\refno=1
\newwrite\rfile
\def\defref{$^{{\hbox{\rfont [\the\refno]}}}$\nref}
\def\nref#1{\xdef#1{\the\refno}\writedef{#1\leftbracket#1}%
\ifnum\refno=1\immediate\openout\rfile=refs.tmp\fi
\global\advance\refno by1\chardef\wfile=\rfile\immediate
\write\rfile{\noexpand\item{#1\ }\reflabeL{#1\hskip.31in}\pctsign}\findarg}
\def\findarg#1#{\begingroup\obeylines\newlinechar=`\^^M\pass@rg}
{\obeylines\gdef\pass@rg#1{\writ@line\relax #1^^M\hbox{}^^M}%
\gdef\writ@line#1^^M{\expandafter\toks0\expandafter{\striprel@x #1}%
\edef\next{\the\toks0}\ifx\next\em@rk\let\next=\endgroup\else\ifx\next\empty%
\else\immediate\write\wfile{\the\toks0}\fi\let\next=\writ@line\fi\next\relax}}
\def\striprel@x#1{} \def\em@rk{\hbox{}} 
\def\lref{\begingroup\obeylines\lr@f}
\def\lr@f#1#2{\gdef#1{\defref#1{#2}}\endgroup\unskip}
\newwrite\lfile
{\escapechar-1\xdef\pctsign{\string\%}\xdef\leftbracket{\string\{}
\xdef\rightbracket{\string\}}}

\def\writestop{\def\writestoppt{\immediate\write\lfile{\string\p
ageno%
\the\pageno\string\startrefs\leftbracket\the\refno\rightbracket%
\string\def\string\secsym\leftbracket\secsym\rightbracket%
\string\secno\the\secno\string\meqno\the\meqno}\immediate\closeout\lfile}}
\def\writestoppt{}\def\writedef#1{}
\catcode`\@=12 
\def\hboX#1{{\sevenrm{\hbox{#1}}}}

\line{\hfill DAMTP 96/64}
\bigskip
\centerline{\fourteenbf SIMPLE PHYSICAL APPROACH}
\smallskip
\centerline{\fourteenbf TO THERMAL CUTTING RULES}
\bigskip
\centerline{P V Landshoff$^*$}
\centerline{DAMTP, University of Cambridge}
\footnote{}{$^*$ email pvl@damtp.cam.ac.uk}
\vskip 15truemm

{\bf Abstract}

A first-principles derivation is given of the imaginary part of a 
Green's function in real-time thermal field theory. The analysis and its
conclusions are simpler than in the usual circled-vertex formalism.
The relationship to Cutkosky-like cutting rules is explained.
\vskip 15truemm
In a recent paper, Bedaque, Das and Naik\defref\bedaque{
P F Bedaque, A Das and S Naik, hep-ph/9603325
} have discussed some cancellations among diagrams that
contribute to the
imaginary parts of finite-temperature amplitudes. A purpose of this note
is to point out that any such  cancellations occur because the usual treatment
of the imaginary parts, which starts in $x$ space and introduces
circled vertices\defref\cutting{
M Veltman, {\it Diagrammatica}, Cambridge University Press (1994)
}, is unnecessarily complicated. A momentum-space 
treatment, which starts from first principles\defref\jct{
P V Landshoff and J C Taylor, Nuclear Physics B430 (1994) 683
},
and stays close to the physics leads to the answer much more simply and 
directly.

Suppose, for definiteness, that we want to calculate\defref\weldon{
L D McLerran and T Toimela, Physical Review D31 (1985) 545\h
H A Weldon, Physical Review D42 (1990) 2384
}  dilepton production from a plasma at temperature $1/\beta$.
The plasma is defined to be composed of quarks and gluons only, with any
leptons or photons that may be produced immediately escaping from it.
The initial density matrix is
$$
\rho _0 =Z^{-1}{\cal P}\exp (-\beta H^{\hboX{}})
\eqno(1a)
$$
where $H^{\hboX{}}$ is the hadronic part of the Heisenberg-picture
Hamiltonian and ${\cal P}$ is a projection operator which removes any states
that do not contain just quarks and gluons. We may choose to express ${\cal P}$
in terms of a complete orthonormal set of plasma states $|i\hbox{ in}\rangle$, 
so that
$$
\rho _0 =Z^{-1}\sum _i |i\hbox{ in}\rangle\langle i\hbox{ in}|\;
e^{-\beta H}
\eqno(1b)
$$
The partition function $Z$ is defined in the usual way, so as to make
tr$\,\rho _0 =1$. The probability of emission of a lepton pair 
of total momentum $q$ is then calculated from
$$
W_{\mu\nu}(q)=\int d^4x\,e^{iq.x}\sum _f \langle f\hbox{ out}|J_{\mu}(x)\,
\rho _0\, J_{\nu}(0)| f\hbox{ out}\rangle
\eqno(2)
$$
where $J(x)$ is the hadronic part of the Heisenberg-picture
electromagnetic current and $| f\hbox{ out}\rangle$
are a complete set of final plasma states.

The usual zero-temperature perturbation theory applies to the matrix
elements that arise in the sum (2). Introduce first an interaction picture
that coincides with the Heisenberg picture at time $t_0$:
$$
J(x)=U^{t_0}(t_0,x^0)\;J^{t_0}(x)\;U^{t_0}(x^0,t_0)
\eqno(3)
$$
where
$$
U^{t_0}(t_2,t_1)=\Lambda(t_2)\Lambda ^{-1}(t_1)=
T\exp\left (-i\int _{t_1}^{t_2}dt\int d^3x \,
H^{t_0}_{\hboX{INT}}(t,{\bf x})\right )$$$$
\Lambda (t)=e^{i(t-t_0)H_0^{t_0}}e^{i(t-t_0)H}
\eqno(4)
$$
Here $T$ denotes ordinary time-ordering and
$H^{t_0}_{\hboX{INT}}$ is the interaction Hamiltonian density written
as a functional of interaction-picture fields and canonical momenta.
We have
$$
e^{-\beta H}=e^{-\beta H_0^{t_0}}\;U^{t_0}(t_0-i\beta,t_0)
\eqno(5)
$$
We take the limit $t_0\to -\infty$, so that the interaction-picture
fields become the $in$ fields. We use also
$$
| f\hbox{ out}\rangle =U^{\hboX{in}}(-\infty ,\infty )| f\hbox{ in}\rangle
\eqno(6)
$$
Then
$$
W_{\mu\nu}(q)=Z^{-1}\int d^4x\, e^{iq.x}\sum_{i,f}\langle f\hbox{ in}|
U^{\hboX{in}}(\infty ,x^0)\;J^{\hboX{in}}_{\mu}(x)\;
 U^{\hboX{in}}(x^0,-\infty ) |i\hbox{ in}\rangle~~~~~~~~~~~~~~~~~~~~~~~~~~~~~~~~
$$$$~~~~~~~~~~~~~~~~~~~~~~~~~~~~~~
\langle i\hbox{ in}|e^{-\beta H^{\hboX{in}}_0}
U^{\hboX{in}}(-\infty -i\beta ,0)\;J^{\hboX{in}}_{\nu}(0)\;
 U^{\hboX{in}}(0,\infty ) |f\hbox{ in}\rangle
\eqno(7)
$$
By using completeness for the states $f$ we obtain:
$$
Z^{-1}\int d^4x\,e^{iq.x}\sum_{i}~~~~~~~~~~~~~~~~~ ~~~~~~~~~~~~~~~~~~~~~~
~~~~~~~~~~~~~~~~~~~~~~~~~~~~~~~~~~~~~~~~~~~~~~~~~~~~~~~~~~~~~~~~~~$$$$
 ~~~~~~~~\langle i\hbox{ in}|e^{-\beta H_0^{\hboX{in}}}
U^{\hboX{in}}(-\infty -i\beta ,0)\;J^{\hboX{in}}_{\nu}(0)\; U^{\hboX{in}}(0,x^0)
J^{\hboX{in}}_{\mu}(x)\; U^{\hboX{in}}(x^0,-\infty )|i\hbox{ in}\rangle
\eqno(8a)
$$
We recognise this\defref\kapusta{
J I Kapusta and P V Landshoff, J Phys G15 (1989) 267
}
as the standard perturbation-theory expression
for
$$
G^{12}_{\mu\nu}(q)=\int d^4x\, e^{iq.x}\tr\big (\rho _0 J_{\nu}(0)J_{\mu}(x)\big )
\eqno(9a)
$$
with the Keldysh contour\defref\keldysh{
L V Keldysh, Sov Phys JETP 20 (1965) 1018
}, which in the complex $t$ plane runs along the real axis from $-\infty$
to $+\infty$, back along the real axis to $-\infty$, and then down to
$-\infty -i\beta$.
We may obtain an alternative expression by
using instead completeness of the states $i$ in (6b), and the fact that 
$H_0^{\hboX{in}}$ is the time-translation operator for the
$in$ fields. We find, instead of (9a),
$$
e^{-\beta q^0}G_{\mu\nu}^{21}(q)=e^{-\beta q^0}\int d^4x\, e^{iq.x}\tr\big (\rho _0 J_{\mu}(x)J_{\nu}(0)\big )
\eqno(9b)
$$ 
For the conserved hermitean electromagnetic current,
$G_{\mu\nu}^{21}(q)$ is symmetric
in $\mu,\nu$ and is real.
Of course, the
equality of (9a) and (9b) is nothing but the standard
relation\ref{\kapusta}
$$ G^{12}(t,{\bf x})=G^{21}(t-i\beta ,{\bf x}) 
\eqno(9c)
$$ 
and one of its consequences is that 
$$
G^{11}_{\mu\nu}(q)=\int d^4x\, e^{iq.x}\tr\big (\rho _0 TJ_{\mu}(x)J_{\nu}(0)\big )
\eqno(10a)
$$
satisfies
$$
\hbox{Im }G^{11}_{\mu\nu}(q)=\half (e^{\beta q^0}+1)G^{12}_{\mu\nu}(q)
\eqno(10b)
$$
We shall find that there are simple generalised Feynman rules to calculate
$G^{12}_{\mu\nu}(q)$, and that they may be recast into a formalism resembling
the zero-temperature Cutkosky formula\defref\elop{
R J Eden, P V Landshoff, D I Olive and J C Polkinghorne, {\it The analytic
$S$-matrix}, Cambridge University Press (1966)
}, though with important differences\defref\kobes{
R L Kobes and G W Semenoff, Nuclear Physics B272 (1986) 329
}.

To derive these, we start with the expression (8a) for $G^{12}_{\mu\nu}(q)$.
It is well-known\defref\lebellac{
M Le Bellac and H Mabilat, Nice preprint INLN 95/38, to appear in Physics
Letters B;\h
T S Evans and A C Pearson, Physical Review D52 (1995) 4652;\h
F G\'elis, hep-ph/9412347
} that one may replace the argument $(-\infty-i\beta )$ in the first $U$
with $-\infty$, so long as one does the same in the calculation of $Z$.
That is, in the original definition (1a) of $\rho _0$ we may replace
$H^{\hboX{}}$ with $H_0^{\hboX{}}$. Of course, the full Hamiltonian
is still used to describe how the system evolves away from its initial
density matrix $\rho _0$. 

The physical reason that allows this modification of the initial
density matrix is as follows. The photon emission rate  which 
we wish to calculate is constant in time and
depends only on the density matrix at the given time. If 
we change the density matrix in the remote past, by multiplying
the interaction by the usual adiabatic switching factor $e^{-\e |t|}$
used in ordinary scattering theory,
this will not affect  the photon emission rate at $t=0$.
In practice, we calculate the total emission over a very long
time interval $T$, and then divide by $T$ by dropping a factor $\delta (0)$.
For most of the time interval $T$ the switching factor has little
effect, and its presence just  changes the result by a term of
order $\e T$ and so it is unimportant.

With this change,
$$
G^{12}_{\mu\nu}(q)= Z^{-1}\int d^4x\, e^{iq.x}\sum_{i,f}e^{-\beta E_i}
\langle i\hbox{ in}| U^{\hboX{in}}(-\infty ,0)\;
J^{\hboX{in}}_{\mu}(0)\; U^{ \hboX{in}}(0,\infty )
|f\hbox{ in}\rangle~~~~~~~~~~~~~~~~~~~~~~~~~~~~~~~~~~~
$$$$~~~~~~~~~~~~~~~~~~~~~~~~~~~~~~\langle f\hbox{ in}|
U^{\hboX{in}}(\infty ,x^0)\;
J^{\hboX{in}}_{\nu}(x)\; U^{\hboX{in}}(x^0,-\infty )
|i\hbox{ in}\rangle
\eqno(11)
$$

The double sum (11) may be calculated from generalised Feynman diagrams
that use the Keldysh-contour $2\times 2$ matrix propagator. In the case
of a scalar field this matrix propagator is
$$
i{\bf D}(k)=\left (\matrix{iD_F(k) & 2\pi\delta ^{-}(k^2-m^2)\cr
                        2\pi\delta ^{+}(k^2-m^2)& (iD_F(k))^*\cr}\right )\;+\;
2\pi\delta (k^2-m^2)\,n(k^0)\left (\matrix{ 1 & 1\cr
                                      1 & 1 \cr}\right )
\eqno(12a)
$$
where $D_F$ is the zero-temperature Feynman propagator and $n$ is
the Bose distribution
$$
n(\o )={1\over e^{\beta |\o |}-1}
\eqno(12b)
$$
There is a similar matrix propagator for the spin-$\half$ case. For gauge
fields, we use the formalism\defref\rebhan{
P V Landshoff and A Rebhan, Nuclear Physics B383 (1992) 607
} in which only the two physical degrees of freedom of the gauge field
have a thermal term; the other two, and the Faddeev-Popov ghost field,
have only the first matrix in (12a). However, for simplicity pretend
first that the fields of the quarks and gluons that make up the
plasma are scalar.

To derive the diagrammatic expansion of (11), expand the $U$'s as in (4)
and apply the zero-temperature Wick theorem to each matrix element
in the double sum.
Each product of operators $UJU$ in the second matrix element
is written as a sum of products of zero-temperature Feynman 
propagators $\langle 0|T\phi\phi|0\rangle$
and normal-ordered creation and annihilation operators.
For the first matrix element we instead have 
$\langle 0|\bar T\phi\phi|0\rangle$ and therefore 
complex conjugates of propagators.
In order to evaluate the matrix elements of the normal 
products, we
must introduce the finite volume $V$ of the plasma, and a discrete
spectrum for the fields:
$$
\phi (x)=\sum _r {1\over\sqrt{2\o _r V}}a_r \exp ({-i\o _rx^0+i\k _r.\x})
\;+\;\hbox{h c}
\eqno(13a)
$$
with
$$
[a_r,a^{\dag}_s]=\delta _{rs}
\eqno(13b)
$$
The states are then labelled by the occupation numbers $n_r$ of the various
single-particle modes $r$.  In the $V\to\infty$ continuum limit
$$
\sum _r\to {V\over (2\pi )^3}\int d^3k$$$$
\phi (x)\to\int{d^3k\over (2\pi )^3}{1\over 2k^0}a(\k )e^{-ik.x}\;+\;\hbox{h c}
$$$$
[a(\k ),a^{\dag}(\k ')]=(2\pi )^3 \;2k^0\;\delta ^{(3)}(\k -\k ')
\eqno(14)
$$
but we do not take this limit until we have used the standard relations
$$
a_r|n_1,n_2\dots\rangle=\sqrt{n_r}|n_1,n_2\dots n_r-1\dots\rangle ,\quad\quad
a^{\dag}_r|n_1,n_2\dots\rangle=\sqrt{n_r+1}|n_1,n_2\dots n_r+1\dots\rangle
\eqno(15)
$$
In order to give a non-zero contribution, the indices $r$ on 
the operators $a$ and $a^{\dag}$ in (11) 
must be equal in pairs. The possible pairings
are

\midinsert{\leftskip 9truemm
\item{($i$)} the index on one of the operators $a$ in the normal product
in the second matrix element matches that on one of the operators $a^{\dag}$
in the anti-normal product in the first matrix element
\item{($ii$)} the index on one of the operators $a^{\dag}$ in the 
second matrix element matches that on one of the operators $a$
in the first matrix element
\item{($iii$)} the index on one of the operators $a$ in the first matrix 
element matches that on one of the operators $a^{\dag}$ in the same matrix
element
\item{($iv$)} the index on one of the operators $a$ in the second matrix 
element matches that on one of the operators $a^{\dag}$ in the same matrix
element
\item{}
}

\endinsert

We need not consider the case where three or more indices
are equal to each other; this would reduce the number of indices left 
to be summed and so\defref\landau{
E M Lifshitz and L P Pitaevskii, in Landau-Lifshitz volume 9 section 13
and volume 10 section 93 (Pergamon)
}, 
according to (14), give fewer powers of $V$ when we go to the continuum limit.
(For a system in thermal equilibrium these nonleading powers actually 
cancel\defref\steer{
T S Evans and D A Steer, hep-ph/9601268 
}, though this is not true more generally.)
For each pair of equal indices $r$,  
we use (15) to replace the corresponding two 
operators with $n_r$ in the cases ($i$),($iii$),($iv$), and $n_r+1$ in case ($ii$).
We then  sum over $n_r$
using
$$
{\left (\sum _{n=0}^{\infty}ne^{-\beta n\o }\right )\over
{\left (\sum _{n=0}^{\infty}e^{-\beta n\o }\right )}} =n(\o )
\eqno(16)
$$
where $n(\o )$ is again the Bose distribution.

\midinsert
\centerline{{\epsfxsize=35truemm\epsfbox{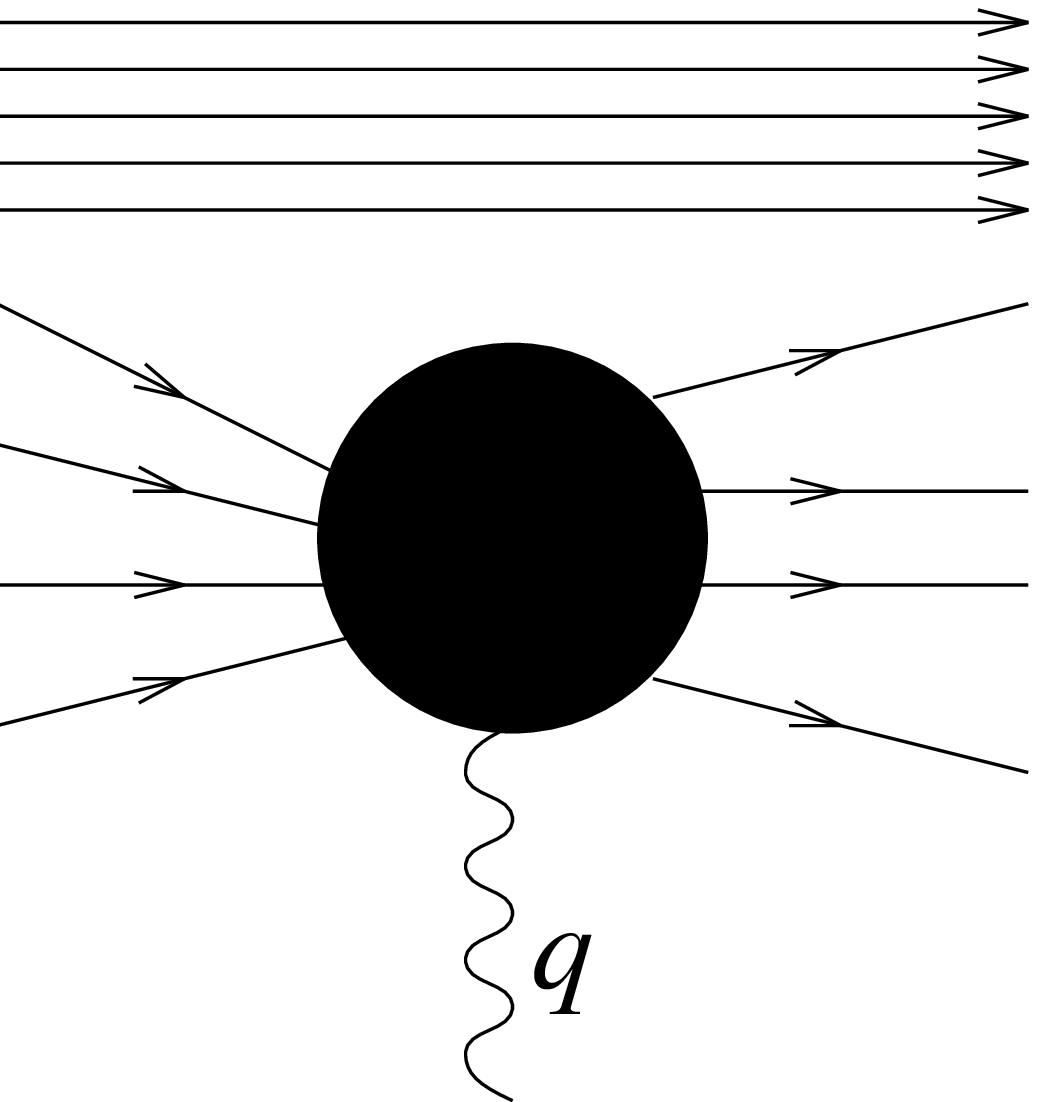}}~~~~~~~~~~~
{\epsfxsize=53truemm\epsfbox{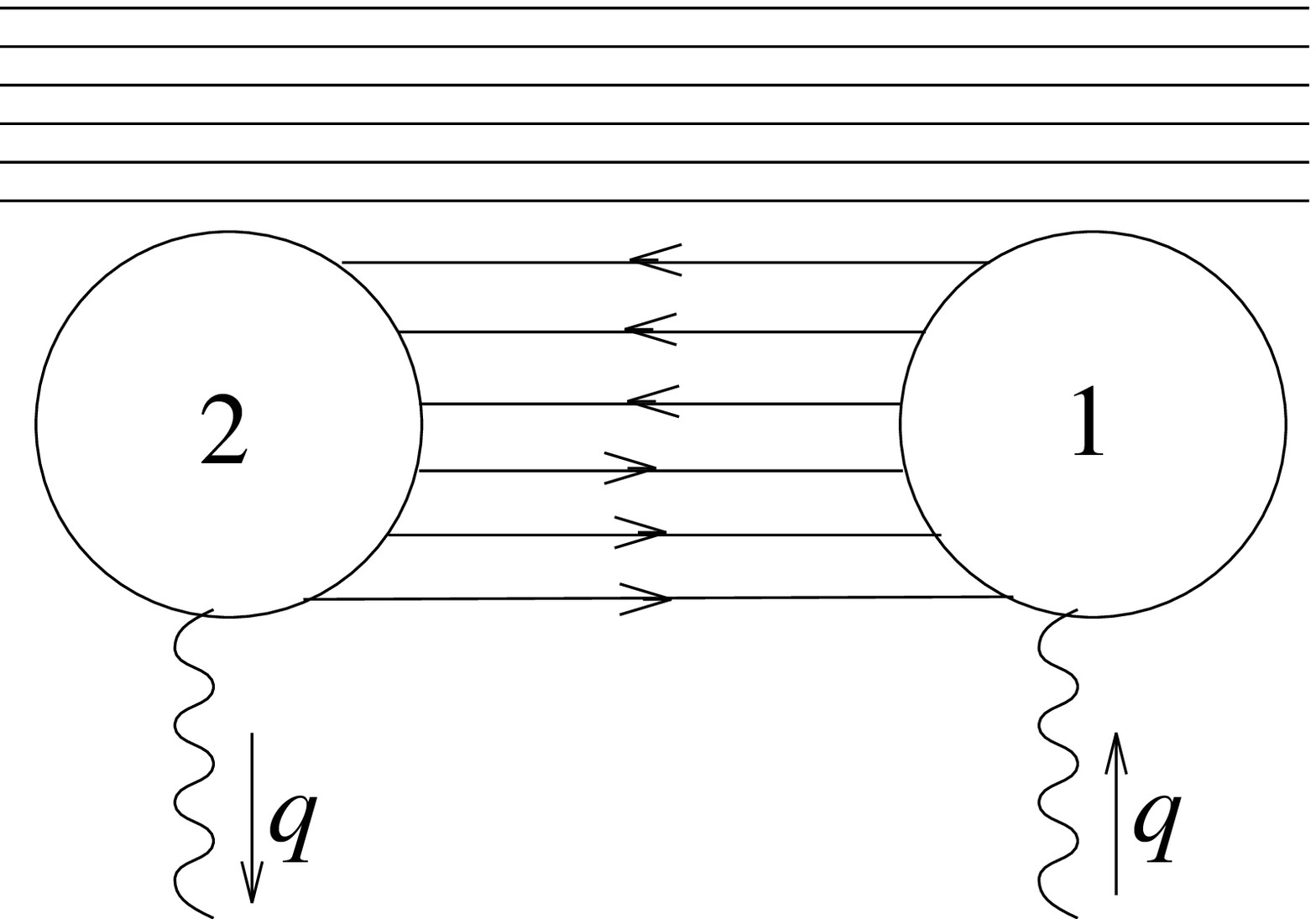}}}\hfill\break
\vskip -5truemm
\centerline{\hfill (a)\hfill (b)\hfill}
\vskip 5 truemm
{\fig Figure 1: (a) a matrix element that contributes to (11), with
(b) its square summed over the momenta. The lines at the top of the
diagrams are the spectator particles in the heat bath.}
\endinsert

\topinsert
\centerline{{\epsfxsize=75truemm\epsfbox{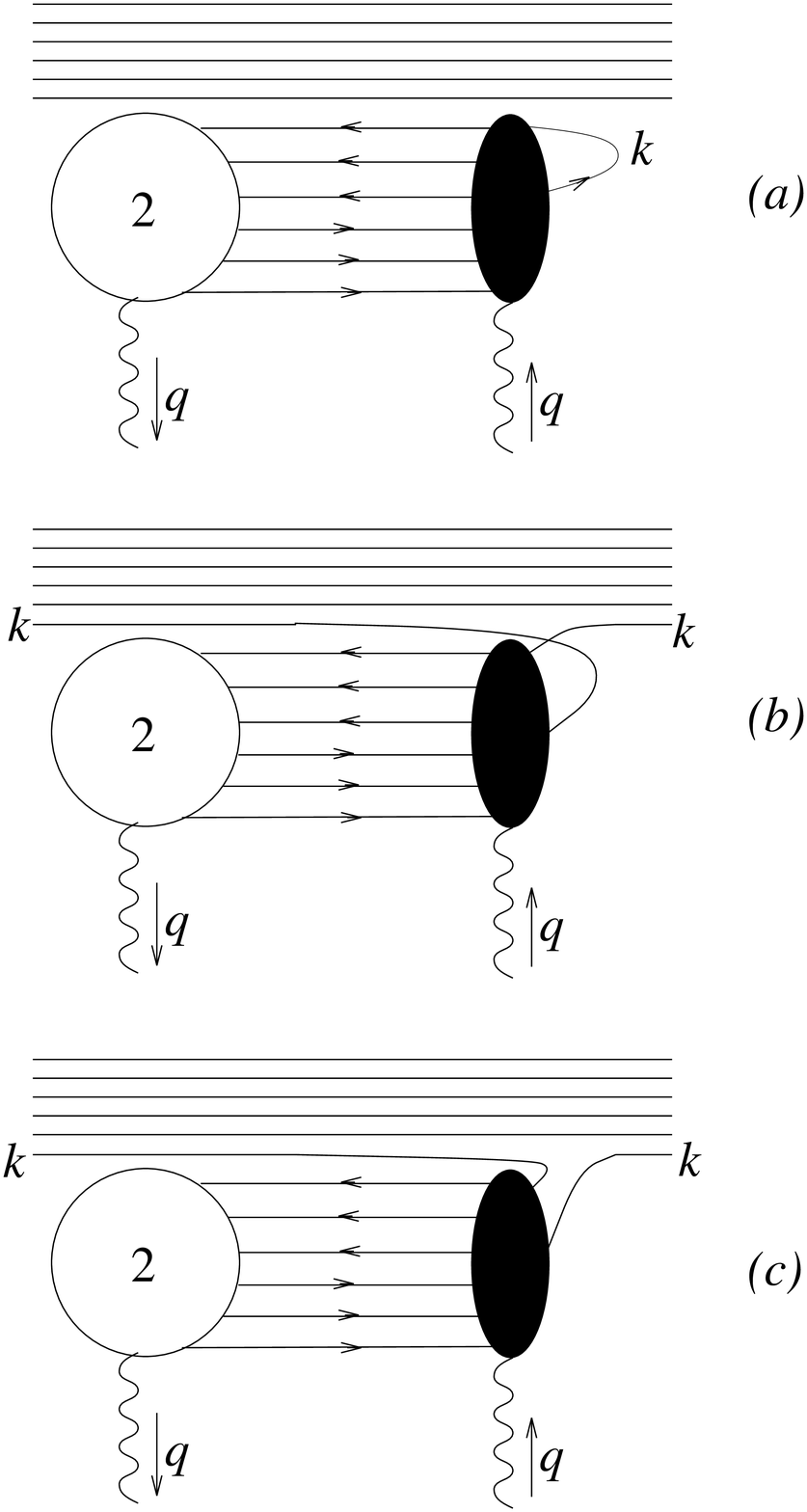}}}\hfill\break
{\fig Figure 2: (a) The diagram of figure 1b with one of the internal
lines $k$ shown explicitly; in (b) and (c) this line instead is a heat-bath
particle.}
\endinsert

We then take the continuum limit and represent the results diagrammatically. 
Consider first the contribution from cases ($i$) and ($ii$) only. 
In (11) we have a matrix element, shown in figure 1a, which has to be squared
and integrations applied to the momenta on the incoming and outgoing lines. 
The lines at the top of the diagram are the spectator particles of the heat
bath. 
In the squared matrix element the
incoming lines, those on the left in figure 1a, correspond to case ($i$); 
with each of them
is associated $2\pi\delta ^+(k^2-m^2)n(k^0)$. The outgoing lines on the right
correspond to case ($ii$); with them is associated 
$2\pi\delta ^+(k^2-m^2)(n(k^0)+1)$. 
The result of squaring figure 1a and performing the momentum integrations
may be depicted
as in figure 1b, where 2 labels the first matrix element in (11)
and 1 the second. To agree with what has been said,
the lines joining them correspond to the $\{21\}$
element of the thermal propagator matrix $i{\bf D}$ of
(12a); the direction of the arrow 
denotes whether the energy flow from 2 to 1 is positive or negative.
The matrix element 1 is an ordinary zero-temperature matrix element;
its internal lines are zero-temperature Feynman propagators $iD_F$. 
The matrix element 2 is its complex conjugate, so its internal lines are
rather $(iD_F)^*$.

\topinsert
\centerline{{\epsfxsize=73truemm\epsfbox{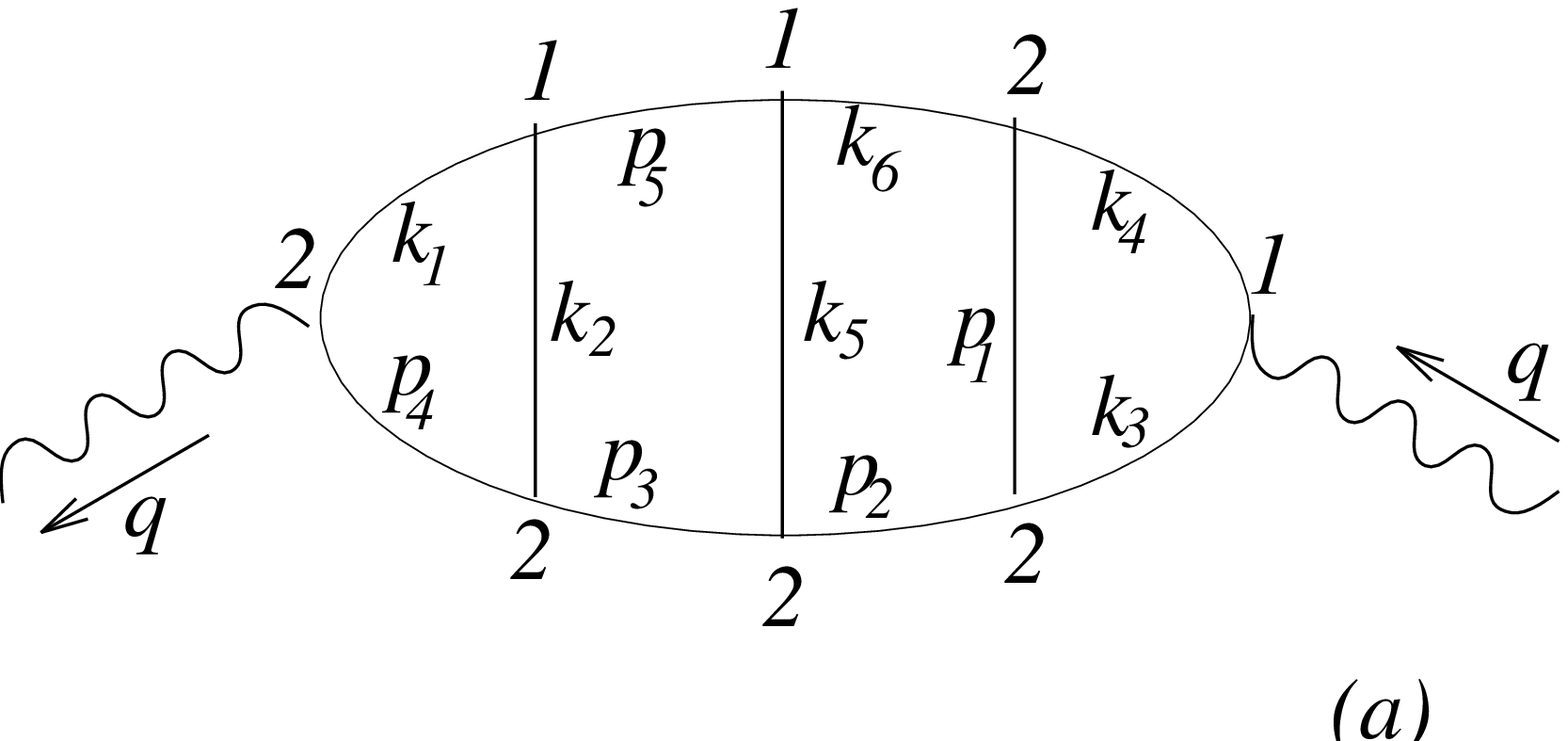}}}\hfill\break
\centerline{{\epsfxsize=73truemm\epsfbox{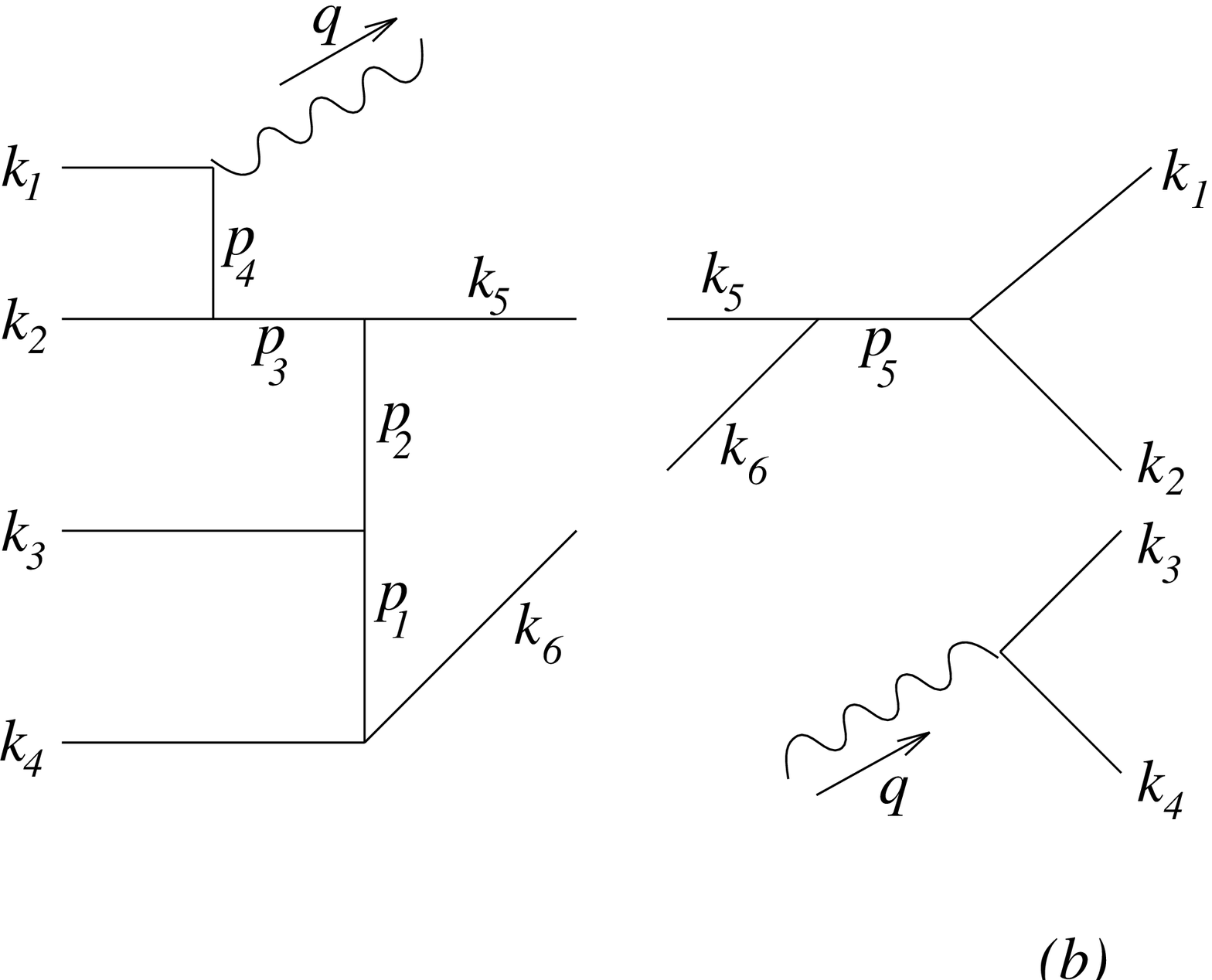}}}\hfill\bigskip
{\fig Figure 3: (a) an ``uncuttable graph'', with an example
(b) of a pair of interfering physical processes to which it corresponds}
\endinsert

In figure 2a we expose one such internal line $k$ in the matrix element 1.
Figures 2b and 2c show diagrams corresponding to case ($iii$). The internal
line $k$ is replaced with two external lines of the same momentum. In
the matrix element 1, this corresponds to a heat-bath particle $k$
undergoing forward scattering and returning to the heat bath with the
same momentum. This matrix element is multiplied by a corresponding matrix 
element 2 in which the heat-bath particle $k$ is a spectator. Figure 2b
may be obtained from figure 2a by replacing $iD_F(k)$ with
$2\pi\delta ^+(k^2-m^2)n(k^0)$, and for figure 2c we instead need
$2\pi\delta ^-(k^2-m^2)n(k^0)$. If we add the three figures together, $iD_F(k)$
becomes just the $\{11\}$ element 
of the thermal propagator matrix $i{\bf D}$. Exactly similar arguments apply
to case ($iv$) and 
the matrix element 2, where instead we arrive at the $\{22\}$ element
of $i{\bf D}$. So we may omit graphs where there
is interference with spectator particles in the heat bath, and instead
use thermal propagators on internal lines of the matrix elements 1 and 2.
Then we need no longer explicitly
draw the heat-bath spectators in our diagrams.

Notice that the matrix elements 1 and 2 need not both be connected.
This can lead to the ``uncuttable'' diagrams of Kobes and
Semenoff\ref{\kobes}.
An example of such an uncuttable graph is figure 3a. One of the 
physical processes to which it corresponds is shown in figure 3b, where
a connected graph on the left interferes with a disconnected one on the
right.

The prescription, then, is that the incoming-current vertex is of type 1
and the outgoing one of type 2. For the internal vertices we sum over
either possibility. Of course, some of the resulting graphs vanish for
kinematic reasons. An example is figure 4: one of its internal
vertices is a type-1 vertex
connected just to three type-2 vertices. Because the $\{21\}$ element
of $i{\bf D}$ puts the corresponding line on-shell, and -- at least in the case
of nonzero mass -- it
is kinematically impossible to have a three-line vertex where all
three lines are on shell, the graph vanishes.

\midinsert
\centerline{{\epsfxsize=73truemm\epsfbox{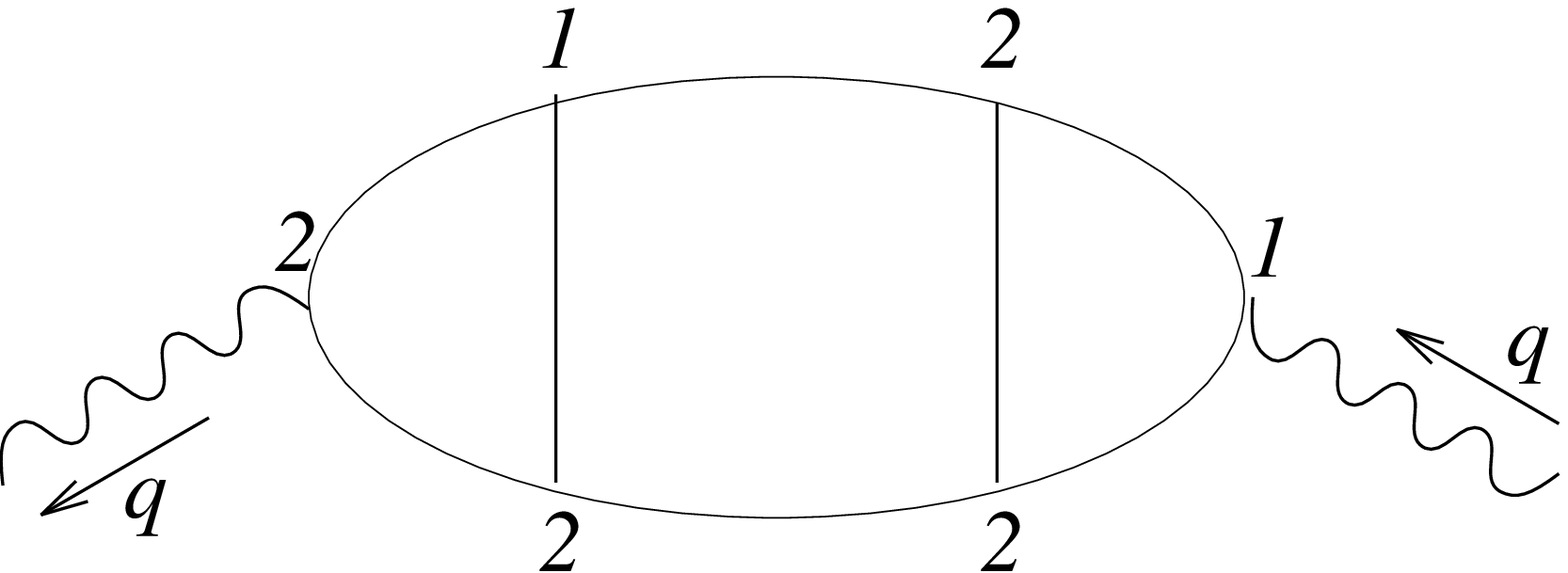}}}\hfill\break
\centerline{\fig Figure 4: a graph that vanishes for kinematic reasons}
\endinsert

The graphs of figure 5 contain ``self-energy'' insertions. Figures 5a and
5b are associated with forward scattering on a heat-bath particle, as
has been explained above, while figure 5c represents either the absorption
of a particle from the heat bath or the emission of an extra one into it.
The graph of figure 5d
vanishes for the kinematic reasons explained
above. These self-energy diagrams are delicate, in that two singular propagators
having the same argument are multiplied together, but by switching off
the interaction in the remote past and future one may show\defref\jacob{
M Jacob and P V Landshoff,  Physics Letters B281 (1992) 114
} that one should handle them in a way similar to zero-temperature
self-energy insertions. The thermal propagator $i{\bf D}(k)$ of (12a)
may be written as\defref\rebhan2{
P V Landshoff and A Rebhan, Nuclear Physics B410 (1993) 23
}
$$
i{\bf D}(k)=\M(k)(i\hat\D(k))\M(k)$$$$
\M(k)=\sqrt{n(k_0)}\left (\matrix{e^{\b |k^0|/2}& e^{-\b k^0/2} \cr
e^{\b k^0/2}& e^{\b |k^0|/2} \cr}\right )$$$$
 i\hat\D(k)=\left ( \matrix{i\D_F(k)&0\cr
0&-i\D_F^*(k)\cr}\right )
\eqno(17a)
$$
where $D_F$ is, as usual, the zero-temperature Feynman propagator.
The matrix self energy $\P (k)$ has the form
$$
\P(k)=\M^{-1}(k)\hat\P(k)\M^{-1}(k)$$$$
\hat\P(k)=\left (\matrix{\Pi(k)&0\cr
                         0&-\Pi^*(k)\cr}\right )
\eqno(17b)
$$

\midinsert
\centerline{{\epsfxsize=11truecm\epsfbox{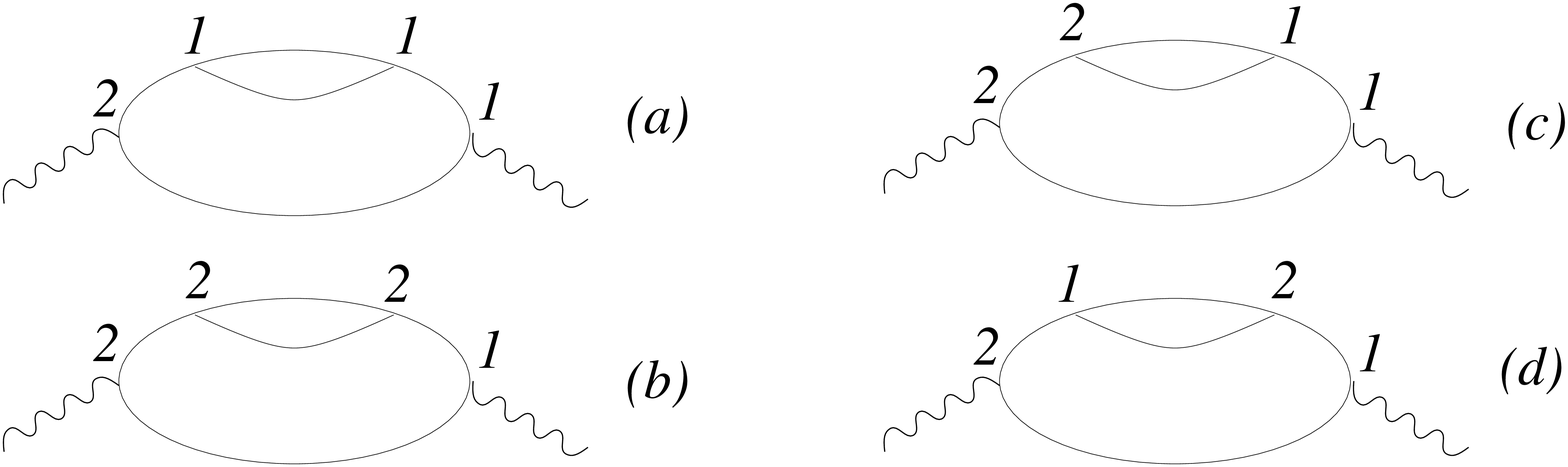}}}\hfill\break
\centerline{\fig Figure 5: graphs with self-energy insertions}
\endinsert

In consequence, the Dyson sum of repeated self-energy insertions
gives the dressed thermal propagator
$$
i{\bf D}'(k)=\M(k)\;(i\hat\D'(k))\;\M(k)$$$$
i\hat\D'(k)=\left (\matrix{i(k^2-m^2-\Pi(k)+i\e)^{-1}&0\cr
                   0&-i(k^2-m^2-\Pi^*(k)-i\e)^{-1}\cr}\right )
\eqno(17c)
$$
To calculate the sum of graphs in figure 5 we need
for the upper line with its insertions
$$
i[\D'(k)]^{21}=i\left [{1\over k^2-m^2-\Pi(k)+i\e}-
{1\over k^2-m^2-\Pi^*(k)-i\e}\right ]n(k^0)
\left (e^{\b k^0}\theta (k^0)+\theta (-k^0)\right )
\eqno(18a)
$$
For values of $k$ for which $\Pi(k)$ is real, this is
$$\eqalign{
i[\D'(k)]^{21}&=2\pi \delta (k^2-m^2-\Pi(k))\; (n(k^0)+\theta (k^0))\cr
&=2\pi Z_{\b}\delta (k^2-m_{\b}^2(k^0))\; (n(k^0)+\theta (k^0))\cr}$$$$
Z_{\b}^{-1}(k^0)=1-{\pd \Pi(k)\over\pd k^2_0}\Big\arrowvert 
_{k^2=m_{\b}^2(k^0)}
\eqno(18b)
$$
where $m_{\b}^2(k^0)$ is the value of $k^2$ for which 
$(k^2-m^2-\Pi(k))$ vanishes. The zero-temperature values of $m_{\beta}$
and $Z_{\b}$ are just the renormalised mass  and the contribution from the
self energy to the charge renormalisation constant. These zero-temperature
values arise just from figures 5a and 5b, not 5c. At nonzero temperature
the additional contributions to $m_{\beta}$ and $Z_{\b}$ may be thought of
as providing extra shifts to the mass and the charge. Usually these are
$k^0$-dependent; they still arise only from figures 5a and 5b because,
as can be checked from (17b), $\Pi ^{21}$ vanishes when $\Pi$ is real.
When $\Pi (k)$ is complex. all three diagrams of figure 5 contribute.

Similar considerations apply to self-energy insertions in the 11 and 22
elements of the thermal propagator ${\bf D}$, which occur within the 1
and 2 matrix elements in figure 1b.

The extension of this treatment to include spin-$\half$ particles is obvious.
In the case of gauge particles, note that the sum over states in (1b) should 
extend only over physical states $i$, since the statistical mechanics from
which it is derived deals only with the possible physical states of the
ensemble. The summation (2) over states is again a summation over
probabilities of achieving any physical state $j$. However, here we may include
also unphysical states, so as to make up a complete set. This is because
a process that begins from a physical state has total probability zero
to end up in an unphysical state: in the canonical quantisation the
Faddeev-Popov ghosts are introduced expressly to achieve 
this\defref\kugo{
T Kugo and I Ojima, Physics Letters 73B (1978) 459
}. Hence we
again arrive at the expression (9a), though with the trace understood
to be restricted to a summation over the physical states $i$. 
In a similar way, if we restrict the states $j$ to be physical, we may extend
the set of states $i$ to make up a complete set, because the ghosts also
ensure that a process that finishes in a physical state has zero total
probability that it began in an unphysical one. So again we may derive (9b).
So  all our diagrammatic analysis is still valid, provided that
any external gauge particles in the graphs are restricted to their
physical degrees of freedom, with no external ghost lines, 
and for internal gauge particles only the
physical degrees of freedom are thermalised, with their unphysical modes and 
the ghosts having just the zero-temperature Feynman propagators\ref{\rebhan}.
Of course, when zero-mass gauge particles are present there are various
infrared divergences. These must cancel in physical quantities,
and the gauge invariance together with unitarity can also cause
other cancellations\ref{{\jct},\jacob}. 

\bigskip\bigskip
{\sl I am grateful to Petr Jizba for arguing with me.\h
This research is supported in part by the EU Programme ``Human Capital
and Mobility", Network ``Physics at High Energy Colliders'', contract
CHRX-CT93-0357 (DG 12 COMA), and by PPARC.}
\vfill\eject
\medskip\immediate\closeout\rfile\writestoppt
\baselineskip=10pt{{\bf References}}\bigskip{\frenchspacing%
\parindent=20pt\escapechar=` \input refs.tmp\bigskip}\nonfrenchspacing
\bye